\documentclass[11pt]{article}
\newcommand{\newsection}[1]{\section{#1}\setcounter{equation}{0}}

\newcounter{newapp}
\usepackage{amsmath, amssymb}
\usepackage{euscript}
\usepackage{amssymb}
\usepackage{euscript}
\usepackage{epsfig,bbm}

\def\a{{\alpha}}
\def\b{{\beta}}

\def\bra#1{\langle #1 |}
\def\ket#1{|#1 \rangle}
\def\0{\nonumber}

\def\det{{\rm det}}
\def\Det{{\rm Det}}

\def\exp{{\rm exp}}

\newcommand\N{{\cal{N}}}

\newcommand\ee{\end{eqnarray}}      
\newcommand\be{\begin{eqnarray}}
\newcommand\ba{\begin{array}}           
\newcommand\ea{\end{array}}
\newcommand\eeq{\end{equation}}     
\newcommand\beq{\begin{equation}}

\begin{document}
\begin{flushright}
 SISSA 02/05/EP\\ hep-th/0501011

\end{flushright}

\vspace{1.2in}
\begin{center}
{\Large\bf Time--localized projectors in String Field Theory with
$E$--field}
\end{center}
\vspace{0.2in}
\begin{center}
C.Maccaferri\footnote{maccafer@sissa.it},
R.J.Scherer Santos\footnote{scherer@sissa.it}, D.D.Tolla\footnote{tolla@sissa.it}\\
\vspace{2mm}

{\it International School for Advanced Studies (SISSA/ISAS)\\
Via Beirut 2--4, 34014 Trieste, Italy, and INFN, Sezione di
Trieste}\\
\end{center}
\vspace{.75in}
\begin{center}
{\bf Abstract}
\end{center}
We extend the analysis of hep-th/0409063 to the case of a constant
electric field turned on  the worldvolume and on a transverse
direction of a  D--brane. We show that time localization is still
obtained by inverting  the discrete eigenvalues of the lump
solution. The lifetime of the unstable soliton is shown to depend
on two free parameters: the $b$--parameter  and the value of the
electric field. As a by--product, we construct the normalized
diagonal basis of the star algebra in $B_{\mu\nu}$--field
background.
~\\
\vspace{5cm} \pagebreak

\newsection{Introduction}

The decay of non--BPS D--branes through open string Tachyon
Condensation is an important  phenomenon  as far as time
dependence in String Theory is concerned. It is clear by now that
the study of Tachyon Condensation can  be naturally implemented in
the framework of Open (Super) String Field Theory, \cite{Witten}.
The theory can be formulated on an unstable D--brane, and evidence
for a stable closed string vacuum (the tachyon condensate) has
been given in great abundance (see \cite{Ohmori1, arefeva, WTBZ,
Senreview} for review and references therein). However, at the
moment,
 no analytical explicit solution has been given to the
classical equation of motion of OSFT, with the appropriate
characteristics of the tachyon vacuum. Even more subtle is the
problem of finding time dependent solutions interpolating from the
unstable D--brane to the pure closed string background arising
after tachyon condensation. Some attempts towards obtaining these
solutions in Cubic Open String Field Theory were made in
\cite{Kluson}. In \cite{GrossE} it was shown that the time
coordinate given by the center of mass in the time direction fails
to be a causal choice of time for which a consistent initial
values problem can be defined, in this regard it was proposed that
a causal time coordinate is the midpoint component of the
light--cone time. In \cite{Erler} it was given some numerical
evidence in (modified) level truncation that the midpoint
light--cone time dependent solutions avoid exploding oscillations
and converge to some finite value of the zero momentum tachyon.
Despite this progress, it is still a challenge to find  an exact
analytic method to attack classical OSFT.

In this regard, Vacuum String Field Theory, \cite{Ras}, although
formulated in a singular way with respect to a regular worldsheet
geometry (it is supposedly obtained from Witten's OSFT by a
singular field redefinition which contracts the open string to its
midpoint, hence shrinking worldsheet boundaries to zero size
holes), \cite{GRSZ1}, is remarkably successful in describing
analytically classical solutions (open string vacua), which
correspond to idempotents of the matter star algebra, the ghost
part of the solution being universal for all kinds of  open string
backgrounds, because of the universal
 nature of the kinetic operator. VSFT is conjectured to represent
OSFT around the to--be--found tachyon vacuum and has the great
advantage of having a kinetic operator which is simply a
$c$--midpoint insertion. Of course one has to pay the cost of a
such drastic simplification in terms of many ambiguous quantities
that arise while computing observables, these ambiguities being
related to midpoint singularities, which need some  regularization \cite{MT, GRSZ1, Oka,  BMP1, BMP2}.
Nevertheless, it is still very attractive to consider star algebra
projectors as describing, at least in the leading order of some
consistent regularization scheme, static and time dependent open
string vacua such as D--branes and their classical decay.

 Driven by these
considerations, the authors of \cite{BMST} have shown that it is
possible to get time localized solutions of VSFT by  taking an
euclidean lump solution on a transverse direction (the euclidean
time) and simply inverting the discrete eigenvalues of the lump
Neumann matrix. This solution, preserving the same euclidean
action as the conventional lump \footnote{We stress that in
presence of time dependent backgrounds, one cannot anymore consider
the value of the classical action mod volume as the space
averaged energy}, has the remarkable feature of being localized in
the time coordinate identified by the twist even discrete
eigenvector of the Neumann matrices and, what is more important,
of being localized in the center of mass time coordinate for every
value of the free parameter $b$. Moreover, it was found that time
dependence disappears when $b\to\infty$ (the solution becoming the
zero momentum sliver state) and when $b\to 0$ (the solution
becoming the 0 string field, i.e. the ``stable'' vacuum of VSFT).
This leads the authors of \cite{BMST} to propose that, at least in
these singular limits, the $b$--parameter should be related to the
$\tilde\lambda$ parameter of Sen's Rolling Tachyon BCFT.\footnote{Another approach in obtaining
time dependent solutions in VSFT can be found in \cite{hata}}

In the present work we are going to study the corresponding time
dependent solution in the presence of a constant $E$--field
background on a longitudinal or transverse direction. We obtain
$E$--field physics by first going to a euclidean signature with
{\it imaginary} $B$--field, $B=iE$, and then inverse Wick
rotating, in  the same way as  \cite{Mukho} for what concerns
the effective target space and BCFT analysis.

One of the main differences with respect to \cite{BMST} is that
when the $E$--field reaches its critical value
$E_c=\frac{1}{2\pi\alpha'}$, the (center of mass) time dependence
is  lost, regardless of the $b$--parameter, and we get a flat non
zero time profile which, along the lines of \cite{Mukho} should be
interpreted as a static background of fundamental strings,
polarized by the $E$--field. This result persists when $b\to 0$ if
we double scale appropriately $E\to E_c$ with $b\to 0$. A direct
construction of the fundamental string background in VSFT is proposed
in \cite{lorianofund}.

The rest of this paper is organized as follows. In Section 2, we
construct the solution representing the decay of a D25--brane with
an $E$--field turned on along a longitudinal direction and show
that this solution is obtained from  the corresponding solution
without $E$--field, where the flat metric along the direction of
the $E$--field and the transverse euclidean time direction, is
replaced with the open string metric. We show that in this case
the  $E$--field  manifests itself  via the life time of the
D--brane, which is longer than its value without $E$--field and
can be infinite for the $E$--field approaching its critical value,
at which  (open) strings becomes effectively tensionless and
closed strings decouple, \cite{SussSeib}. In Section 3, we
construct a solution which represents the brane  decay with an
$E$--field turned on along a transverse direction and show that,
although the $E$--field coupling to transverse momenta cannot
 be anymore disregarded, the physical situation is the same as  the
longitudinal case, but with localization on the transverse spatial
direction, except at the extreme value $b=\infty$. In Section 4,
we make a summary and discuss our results. In the first two
appendices  we carry out a detailed analysis of the spectroscopy
of the Neumann matrices in $B(E)$--field background and explicitly
diagonalize the 3-string vertex. This analysis fills some gaps of
\cite{Feng}, in particular we  appropriately fix the
normalizations of continuous and discrete eigenvectors and, for
the latter, we give a different expression which, contrary to
\cite{Feng}, reproduces the known results for $B=0$,
\cite{belov1}. In appendix C we study the asymptotic behavior of
Neumann matrices for the relevant limits  $b\to \infty$ and
$b\to0$.

\newsection{Longitudinal $E$--field}

In this section we will analyze the case of switching  the
$E-$field along a tangential direction, i.e., along, say, the world volume of
a $D25-$brane. As explained in \cite{Mukho}, the presence of the $E$--field does
not create non commutativity as the direction in which it is turned on is at zero
momentum.

We use the double Wick rotation, that is we make space--time
euclidean by sending $X^0(\sigma)\to i X^D(\sigma)$; then we
construct  an  {\it unconventional} lump solution, \cite{BMST}, on the
transverse spatial direction $X^D(\sigma)$ and inverse Wick rotate
along it, $X^D(\sigma)\to -i X^0(\sigma)$. Let  $\alpha,\beta=1,D$
be the couple of directions on which the $E$--field is turned on.
Then $E$--field physics is obtained by taking an imaginary
$B$--field
\be
B_{\alpha\beta}=B\epsilon_{\alpha\beta}=iE\epsilon_{\alpha\beta},\quad
E\in \Re
\ee
A localized time dependent solution is easily given by
straightforwardly changing the metric $\eta_{\alpha\beta}$ of the
 solution of \cite{BMST}, with the open string metric $G_{\alpha\beta}$
\be\label{osmetric}
G_{\alpha\beta}&=&(1-(2\pi\alpha' E)^2)\,\delta_{\alpha\beta}\\
G^{\alpha\beta}&=&\frac1{1-(2\pi\alpha'
E)^2}\,\delta^{\alpha\beta}
\ee
Note that, contrary to the case of
a real $B$--field, a critical value shows up for the imaginary
analytic continuation\footnote{In the rest of the paper we will set $\alpha'=1$}
\be
E_c=\frac1{2\pi\alpha'}
\ee
 From now on all
indexes $(\alpha,\beta)$ are raised/lowered with the open string
metric (\ref{osmetric}).

We have then the following commutators
\begin{equation}
[a_{m}^{\alpha},a_{n}^{\beta\dagger}]=
G^{\alpha\beta}\delta_{mn},
~~~~~~~~~~ m,n\geq 1
\label{Bcomm}
\end{equation}
stating that the $a^\alpha$'s are canonically normalized with respect the open string metric (\ref{osmetric})

We recall that, in case of a background $B_{\a\b}$--field, the
three string vertex is deformed to be, \cite{sugino} (see also \cite{kawano})
\be
|V_3\rangle = |V_{3,\perp}\rangle \,
\otimes\,|V_{3,_\|}\rangle\label{split} \ee The factor
$|V_{3,_\|}\rangle$ concerns the directions with no $B$--field and
its expression is the usual one, \cite{GJ1, leclair1, Ohta, tope},
on the other hand  $|V_{3,\perp}\rangle$ deals with the directions
on which the $B$ field is turned on\footnote{Note that in the case
under consideration the symbols $\perp$ and $\|$ do not refer to
perpendicular or
 transverse directions to the brane, but simply indicates directions with
$E$--field turned on ($\perp$) or not ($\|$)} .
\begin{equation}
|V_{3,\perp}\rangle= \int d^{26}p_{(1)}d^{26}p_{(2)}d^{26}p_{(3)}
\delta^{26}(p_{(1)}+p_{(2)}+p_{(3)})\,{\rm exp}(-E')\,
|0,p\rangle_{123}\label{V3}
\end{equation}
The operator in the exponent is given by, \cite{sugino}
\be
E'_\perp &=& \sum_{r,s=1}^3\left(\frac 12 \sum_{m,n\geq 1}
G_{\alpha\beta}a_m^{(r)\alpha\dagger}V_{mn}^{rs}
a_n^{(s)\beta\dagger} +
\sum_{n\geq 1}G_{\alpha\beta}p_{(r)}^{\alpha}V_{0n}^{rs}
a_n^{(s)\beta\dagger}\right.\0\\
&&\left. \quad\quad +\,\frac 12 G_{\alpha\beta}p_{(r)}^{\alpha}V_{00}^{rs}
p_{(s)}^\beta+
\frac i2 \sum_{r<s} p_\alpha^{(r)}\theta^{\alpha\beta}
p_\beta^{(s)}\right)\label{EtoE'}
\ee

Note that the part giving rise to space--time non--commutativity,
$\frac i2 \sum_{r<s} p_\alpha^{(r)}\theta^{\alpha\beta}p_\beta^{(s)}$,
does not contribute due to the zero momentum condition in the 1 spatial direction.\\
Let's first consider the sliver solution at zero momentum along the 1 direction\\
The three string vertex in such a direction takes the form ($p^1=p_1=0$)
\be\label{vert0mom}
|V_3(E,p=0)\rangle&=& |V_3(E=0,p=0)\rangle^{\left(\eta_{11}\to G(E)_{11}\right)}\\
                  &=& \exp\left(\frac12 \sum_{r,s=1}^3G_{11}a^{(r)1\dagger}\cdot V^{rs}\cdot
                   a^{(s)1\dagger}\right)\ket{0}
\ee

This implies that the zero momentum sliver is in this case
\be\label{sli0mom}
|S(E,p=0)\rangle&=& |S(E=0,p=0)\rangle^{\left(\eta_{11}\to G(E)_{11}\right)}\\
                  &=&\N\exp\left(-\frac12 G_{11}a^{1\dagger}\cdot S\cdot
                   a^{1\dagger}\right)\ket{0}
\ee
where the normalization $\N$ and the matrix $S$ are given as usual, \cite{RSZ2},
\be
T&=&CS= \frac 1{2X} (1+X-\sqrt{(1+3X)(1-X)})\label{sliver}\\
\N&=&\sqrt{\det(1-X)(1+T)}
\ee
On the euclidean time direction we need the full 3 string vertex in
oscillator basis. This is given by
\be
|V_{3,\perp}\rangle'= K\, e^{-E'}|\Omega_b\rangle\label{V3'}
\ee
with
\be
&&K= \left(\frac {\sqrt{2\pi b^3}}{3(V_{00}+b/2)^2 }(1-(2\pi E)^2)^\frac12\right)^\frac12,\label{K2}\\
&&E'= \frac 12 \sum_{r,s=1}^3 \sum_{M,N\geq 0} a_M^{(r)D\dagger}
V_{MN}^{'rs} a_N^{(s)D\dagger}G_{DD}\label{E'}
\ee
where $M,N$ denote the couple of indices $\{0,m\}$ and $\{0,n\}$,
respectively, and $D$ is the (euclidean) time direction.
The coefficients $V_{MN}^{'rs}$ are given in Appendix B of \cite{RSZ2}.
In order to have localization in Minkowski time, we need an explosive profile in euclidean time (unconventional lump); this is
explained in  detail in \cite{BMST}
\begin{equation}
\label{rollsliv}
\ket{\check\Lambda'}=\N\exp\left(-\frac12G_{DD}a^{\dagger D} C {\check T}' a^{\dagger D}\right)\ket{\Omega_b}
\end{equation}
where
\begin{equation}
{\check T}'_{NM}=-\int_{-\infty}^\infty dk \,V^{(k)}_N\, V^{(k)}_M
\exp\left(-\frac{\pi|k|}{2}\right)+
\left(V_N^{(\xi)}\,V_M^{(\xi)}+V_N^{(\bar\xi)}\,
V_M^{(\bar\xi)}\right)\exp\,|\eta|\label{T'NM}
\end{equation}
We refer to \cite{BMST} for the exact definition of eigenvalues and eigenvectors of the various Neumann
matrices in the game. We only stress that the Neumann matrix of the unconventional lump has inverted
discrete eigenvalues with respect to the ordinary lump: this, as shown, in \cite{BMST}, is what
guarantees time localization with respect to the center mass and to the time coordinates identified by
the discrete eigenvectors $V_N^{(\bar\xi)},\,V_N^{\xi)}$.

We get a localized time profile by projecting on the coordinates/momenta of the discrete spectrum
\be
\hat x_\eta=\frac{ i}{\sqrt{2}} \left(e_\eta-e_\eta^{\dagger}\right)\\
\hat y_\eta=\frac {i}{\sqrt{2}}
\left(o_\eta-o_\eta^{\dagger}\right)
\ee
where $e_\eta$ / $o_\eta$ are oscillators constructed with the twist even/odd part of the
discrete spectrum eigenvectors $V_N^{(\bar\xi)},\,V_N^{(\xi)}$, see
\cite{BMST}
\be
e_\eta=\sum_{N=0}^{\infty} \frac12 \left(1+(-1)^N\right) V_N^{(\xi)}a_N\\
o_\eta=\sum_{N=0}^{\infty} \frac1{2i} \left(1-(-1)^N\right) V_N^{(\xi)}a_N
\ee
 The profile along these coordinates is given by
(inverse Wick rotation, $(x,y)\to i({\rm x},-{\rm y})$ is
assumed)
\be
\ket{\check\Lambda'({\rm x,y})}=\bra{{\rm
x,y}}\check\Lambda'\rangle=\frac{1}{\pi(1+e^{|\eta|})}
\exp\left(-\frac{e^{|\eta|}-1}{e^{|\eta|}+1} ({\rm
x^2+y^2})\right)|\check\Lambda'_c\rangle
\ee
where $|\check\Lambda'_c\rangle$ contains only continuous spectrum
contributions. This profile is localized on the time coordinate
$\rm x$. Note however that there is no more reference to the
$E$--field in the exponent. In order to see explicitly the
presence of the $E$--field, we need to use the usual {\it open}
string time, i.e. the center of mass.

Therefore we contract our solution with the center of mass euclidean time, $x^D$, and
then inverse Wick rotate it, $x^D\to i x^0$. This is identical to
\cite{BMST}, so we just quote the result, paying attention to use the open string metric (\ref{osmetric})
\be \ket{\Lambda'({\rm
x}_0,y)}&\!=\!&\bra{{\rm x}_0,{\rm y}}\Xi_\eta\rangle=
\sqrt{\frac{2}{b\pi}}\frac{\N}{\sqrt{2\pi(1+e^{|\eta|})}}
\exp\left(\frac{1-e^{|\eta|}}{1+e^{|\eta|}}{\rm y}^2\right)\label{x0yprof}\\
\cdot\frac{1}{\sqrt{1+{\check T}'_{00}}}\!\!&&\!\!
\exp\left(-{\mathcal A}(x^0)^2+
\frac{2i\sqrt{1-(2\pi E)^2}}{\sqrt{b}(1+{\check T}'_{00})}x^0 {\check T}'_{0n}\tilde a_n^\dagger-\frac12
\tilde a_n^\dagger W''_{nm} \tilde a_m^\dagger\right)\ket{0}\0
\ee
The extra coordinate $\rm y$ is given by the twist odd contribution of the discrete spectrum, we need to
project along it in order to have a well defined $b\to\infty$ limit in the oscillator part $W''_{nm}$,
see \cite{BMST}.
The oscillators $\tilde a_n$ are canonically normalized with respect the $\eta$-metric and are given by
\begin{equation}
\tilde a_n=\sqrt{1-(2\pi E)^2}a_n
\end{equation}
 The quantity that give rise to time localization is then
\begin{equation}
{\mathcal A}=-\frac1b\frac{1-{\check T}'_{00}}{1+{\check T}'_{00}}(1-(2\pi E)^2)
\end{equation}
This quantity depends on the free parameter $b$, as well as on the value of the $E$--field, through the
open string metric, used to covariantize the quadratic form in time.
The matrix element ${\check T}'_{00}$ is given in \cite{BMST}

\begin{equation}
{\check T}'_{00}(\eta)=-2\int_0^\infty dk \left(V_0^{(k)}(b(\eta))\right)^2
\exp\left(-\frac{\pi k}{2}\right)+2({V_0^{(\xi)}})^2\exp\,|\eta|,
\end{equation}
 it is a monotonic increasing function of $b$, greater than 1: this is what ensures localization in time as opposed to the
standard lump which is suited for space localization.

The life time of the brane is thus given by \be \Delta
T=\frac{1}{2}\sqrt{\frac{1}{2{\mathcal A}}}=\frac{1}{(1-(2\pi
E)^2)^\frac12}\,\Delta T^{(E=0)} \ee Note that for $E$ going to
the critical value $E_c=\frac{1}{2\pi}$, the lifetime becomes
infinite. In particular we get a completely flat profile. This has
to be traced back to the fact that open strings become effectively
tensionless in this limit, \cite{SussSeib}, so we correctly
recover the result of \cite{BMST}, that the D-brane is stable.
This configuration should correspond to a background of
fundamental strings stretched along the $E$--field direction, with
closed strings completely decoupled.

\newsection{Transverse $E$--field}

 In this section we study the
 time dynamics of a D--brane with transverse $E$--field.
  We will do this in two steps. First we will write down coordinates and momenta
  operators corresponding to the oscillators of the discrete diagonal basis and look at the
  profile of the lump solution with respect to them. Next we will determine the open string
  time profile of the lump solution by projecting it onto
  the center of mass coordinates. Since the solutions with $E$--field are equivalent
   to euclidean  solutions with imaginary $B$--field, before proceeding further, we will first give
   a brief summary of the construction of lump solutions in VSFT with transverse $B$--field.

\subsection{Lump solutions with B field}

The solitonic lump solutions in VSFT in the presence of a constant
transverse $B$ field were determined in \cite{BMS}. The $*$
product is defined as follows
\begin{equation}
_{123}\!\langle V_3|\Psi_1\rangle_1 |\Psi_2\rangle_2 =_3\!\langle
\Psi_1*_m\Psi_2|
\label{starm}
\end{equation}
where the 3-string vertex $V_3$, with a constant $B$ field turned
on along the $24^{th}$ and $25^{th}$ directions (in view of the
D-brane interpretation, these directions are referred to as
transverse), is
\begin{equation}
|V_{3}\rangle=|V_{3,\bot}\rangle\otimes|V_{3,||}\rangle.
\end{equation}
$|V_{3,||}\rangle$ corresponds to the tangential directions while
$|V_{3,\bot}\rangle$ is obtained  from \cite{sugino} by passing to
zero modes oscillator basis and integrating over transverse
momenta, see \cite{BMS}
\begin{equation}
|V_{3,\bot}\rangle={\sqrt{2\pi b^{3}\Delta}\over A^{2}(4a^{2}+3)}{\exp}
\left[{1\over 2}\sum_{r,s=1}^{3}\sum_{N,M\ge0}a_{M}^{(r)\alpha\dagger}{\cal V}_{\alpha\beta,MN}^{rs}a_{N}^{(r)\beta\dagger}\right]
|0\rangle\otimes|\Omega_{b,\theta}\rangle_{123}.
\label{3vert}
\end{equation}
In the following we will set $\alpha, \beta=1,2$
 for simplicity of notation. $|\Omega_{b,\theta}\rangle$ is the vacuum with
respect to the zero mode oscillators
\begin{equation}
a_0^{(r)\alpha} = \frac 12 \sqrt b \hat p^{(r)\alpha}
- i\frac {1}{\sqrt b} \hat x^{(r)\alpha},
~~~~~~~~~~~~
a_0^{(r)\alpha\dagger} = \frac 12 \sqrt b \hat p^{(r)\alpha} +
i\frac {1}{\sqrt b}\hat x^{(r)\alpha}.
\label{osc}
\end{equation}
${\cal V}_{\alpha\beta,MN}^{rs}$ are the Neumann coefficients with
zero modes in a constant $B$ field background, which are symmetric
under simultaneous exchange of all the three pairs of indices and
cyclic in the string label indices ($r,s$) where $r, s=4$ is
identified with $r,s=1$. Moreover  $\Delta=\sqrt{{\rm Det}G}$, $G_{\alpha
\beta}$ being the open string metric along the transverse
directions (\ref{osmetric}). We have also introduced the notations
\begin{equation}\label{aBb}
A=V_{00}+{b\over 2}, ~~~~~~~~~~~~~~ a=-{\pi^{2}\over A}|B|.
\end{equation}

The lump solution is given by
\begin{equation}
|S\rangle=|S_{||}\rangle \otimes {\cal N}{\rm exp}\left( -\frac{1}{2}
\sum_{M,N \ge 0} a^{\alpha \dagger}_{M}{\cal S}_{\alpha \beta,MN}a^{\beta \dagger}_{N}\right)
|0\rangle \otimes |\Omega_{b,\theta}\rangle,
\label{lump}
\end{equation}
where
\begin{equation}
 {\cal N}=\frac{A^{2}(3+4a^{2})}{\sqrt{2\pi b^{3}}(\Det G)^{\frac{1}{4}}} \Det({\cal I}-{\cal X})^{\frac{1}{2}}\Det({\cal I}+{\cal T})^{\frac{1}{2}},
\end{equation}
and
\begin{equation}
{\cal X}=C'{\cal V}^{11},~~~~~~~~{\cal T}=C'{\cal S},~~~~~~~~~C'=(-1)^{N}\delta_{NM}
\end{equation}
In (\ref{lump}) $|S_{||}\rangle$ corresponds to the longitudinal part of the lump solution
 and it is a zero momentum sliver.

In order for (\ref{lump}) to satisfy the projector equation, ${\cal T}$ and ${\cal X}$ should satisfy the relation\footnote{
In this paper we limit ourselves to twist invariant projectors, but our analysis can be straightforwardly generalized
to projectors of the kind \cite{FKM}}
\begin{equation}\label{projeqn}
({\cal T}-1)({\cal X}{\cal T}^{2}-({\cal I}+{\cal X}){\cal T}+{\cal X})=0.
\end{equation}
In the above formulae the $\alpha, \beta, N,M$ indices are implicit.
This equation is solved by ${\cal T}_0$, $1/{\cal T}_0$ and $1$, where
\begin{equation}
{\cal T}_0=\frac{1}{2{\cal X}}\left(1+{\cal X}-\sqrt{(1+3{\cal X})(1-{\cal X})}\right)
\end{equation}
 ${\cal T}=1$
gives the identity state, whereas the first and the second solutions give the lump and the inverse lump, respectively.
In  \cite{BMST} it has been  argued that, although the inverse lump solution was discarded in earlier
works \cite{KP, RSZ2}, because of the bad behaviour of its eigenvalues in the oscillator basis, it is possible to make
sense out of it by considering (\ref{projeqn}) as a relation between eigenvalues relative to twist definite eigenvectors.
In particular, in the diagonal basis, the projector equation  factorizes into the continuous and discrete
 contributions, which separately satisfy  equation (\ref{projeqn}). Therefore, one can just
 invert (for example) the discrete eigenvalues of ${\cal T}$: dangerous $-$ signs under the square root in the
 energy densities of the solution are indeed avoided  by counting the double multiplicity of these eigenvalues, which
 is required by twist invariance. See  Appendix A for
the spectroscopy of ${\cal X}$, and hence of ${\cal T}$.

\subsection{Diagonal Coordinates and Momenta}
In Appendix B $\tau$--twist definite oscillators of the diagonal
basis are introduced. Due to the structure of Neumann coefficients it is natural to define
the twist matrix as $\tau C$, where $\tau=\sigma^3$ acts on space--time indices, see appendices A and B for details.
In the following $C$--parity will be always understood as $\tau C$--parity.
 Now let's define the following coordinates
and momenta operators in terms of the twist even and twist odd parts of the discrete spectrum, (\ref{deftwist})
\begin{equation}
{\hat X}_{\xi_{i}}={i\over\sqrt{2}}(e_{\xi_{i}}-e^{\dagger}_{\xi_{i}})~~~~~~~~~
{\hat Y}_{\xi_{i}}={i\over\sqrt{2}}(o_{\xi_{i}}-o^{\dagger}_{\xi_{i}})
\end{equation}
 which are hermitian by definition and have the following eigenstates
\begin{equation}
|X_{i}\rangle={1\over\sqrt{\pi}}e^{-{1\over 2}X_{i}^{2}-\sqrt{2}iX_{i}e^{\dagger}_{\xi_{i}}+
{1\over 2}e^{\dagger}_{\xi_{i}}e^{\dagger}_{\xi_{i}}}|\Omega_{e_{i}}\rangle
\end{equation}
\begin{equation}
|Y_{i}\rangle={1\over\sqrt{\pi}}e^{-{1\over 2}Y_{i}^{2}-\sqrt{2}iY_{i}o^{\dagger}_{\xi_{i}}+
{1\over 2}o^{\dagger}_{\xi_{i}}o^{\dagger}_{\xi_{i}}}|\Omega_{o_{i}}\rangle.
\end{equation}
We made the assumption that the vacuum factorizes as
\begin{equation}
\ket{0}\otimes|\Omega_{b,\theta}\rangle=\prod_{i=1}^{2}\prod_{k}|\Omega_{i}(k)\rangle\otimes
|\Omega_{e_{i}}\rangle\otimes|\Omega_{o_{i}}\rangle
\end{equation}
where $|\Omega_{i}(k)\rangle$, $|\Omega_{e_{i}}\rangle$ and $|\Omega_{o_{i}}\rangle$ are vacua
with respect to the continuous, the twist even discrete and twist odd discrete oscillators, respectively.

The explicit $(X_{i},Y_{i})$ dependence of the lump state (\ref{lump}) can be obtained
by projecting it onto the eigenstates $|X_{i},Y_{i}\rangle$. After re-writing (\ref{lump}) in terms of
the diagonal basis oscillators and performing the projection (see Appendix B), it follows
$$
\langle X_{i},Y_{i}|S\rangle={1\over \pi^{2}[1+ t_d(\eta_{1})][1+ t_d(\eta_{2})]}{\rm exp}{1\over 2}
\left[{t_d(\eta_{1})-1\over t_d(\eta_{1})+1}(X_{1}^{2}+Y_{1}^{2})\right.$$
\begin{equation}
~~~~~~~~~~~~~~~~~~~~~~~~~~~~~~~~~\left.+{t_d(\eta_{2})-1\over t_d(\eta_{2})+1}(X_{2}^{2}+Y_{2}^{2})\right]
|S\rangle_{c}\otimes|S_{||}\rangle .
\label{lumpd12}
\end{equation}
$|S\rangle_{c}$ is given by (\ref{lumpd}) with only continuous spectrum oscillators and $t_d(\eta_{i})=e^{-|\eta_{i}|}$
are the discrete eigenvalues of ${\cal T}$ corresponding to the eigenvalue $\xi(\eta_{i})$ of the operator $C'{\cal U}$.

In (\ref{lumpd12}) the directions $\alpha, \beta$ are completely mixed. As a matter of fact, it is not apparent at
this stage which of these variables ($X_{i}, Y_{i}$) contain the information about the center of mass time
dependence of the lump. To make this  clear let's recall  the non-diagonal basis oscillators and write
the coordinates and the momenta operators as
\begin{equation}
{\hat X}_{N}^{\alpha}={i\over\sqrt {2}}(a_{N}^{\alpha}-a_{N}^{\alpha\dagger})~~~~~~~~~
{\hat P}_{N}^{\alpha}={1\over\sqrt {2}}(a_{N}^{\alpha}+a_{N}^{\alpha\dagger}).
\end{equation}
 In order to get the relation between these operators and the corresponding diagonal operators
 we have defined above, we need to re-write the diagonal basis oscillators in terms of the non-diagonal ones.
 In doing so, one has to be careful about taking the complex conjugate of the eigenstates,
 as we are dealing with hermitian rather then symmetric matrices. Taking this fact into account
  and using some results of Appendix B, we obtain
\begin{equation}
e_{\xi_{i}}={1\over\sqrt{2}}\sum_{N=0}^{\infty}(V_{N}^{(\xi_{i})\alpha}+V_{N}^{({\bar\xi}_{i})\alpha})
a_{N,\alpha}~~~~~~~~e_{\xi_{i}}^{\dagger}={1\over\sqrt{2}}\sum_{N=0}^{\infty}({\bar V}_{N}^{(\xi_{i})\alpha}+
{\bar V}_{N}^{({\bar\xi}_{i})\alpha})a_{N,\alpha}^{\dagger}
\end{equation}
\begin{equation}
o_{\xi_{i}}={-i\over\sqrt{2}}\sum_{N=0}^{\infty}(V_{N}^{(\xi_{i})\alpha}-V_{N}^{({\bar\xi}_{i})\alpha})a_{N,\alpha}~~~~~~~~
o_{\xi_{i}}^{\dagger}={i\over\sqrt{2}}\sum_{N=0}^{\infty}({\bar V}_{N}^{(\xi_{i})
\alpha}-{\bar V}_{N}^{({\bar\xi}_{i})\alpha})a_{N,\alpha}^{\dagger}\label{twdefo}
\end{equation}
and similar relations for the continuous spectrum oscillators.
Hence, the diagonal coordinates and momenta can be written as
\begin{equation}
{\hat X}_{\xi_{i}}=\sqrt{2}\sum_{N=0}^{\infty}V_{2N}^{\xi_{i},1}{\hat X}_{2N}^{1}+V_{2N+1}^{\xi_{i},2}{\hat P}_{2N+1}^{2}
\end{equation}
\begin{equation}
{\hat Y}_{\xi_{i}}=\sqrt{2}\sum_{N=0}^{\infty}V_{2N+1}^{\xi_{i},1}{\hat P}_{2N+1}^{1}-iV_{2N}^{\xi_{i},2}{\hat X}_{2N}^{2}
\end{equation}

 Now, to make the center of mass time dependence of the solution explicit, we need to
 extract  the zero modes from these operators. Let's write the zero mode coordinate
  and momentum operators by introducing the b parameter as
\begin{equation}
{\hat X}_{0}^{\alpha}={i\over\sqrt b}(a_{0}^{\alpha}-a_{0}^{\alpha\dagger})~~~~~~~~~
{\hat P}_{0}^{\alpha}={\sqrt{b}\over 2}(a_{0}^{\alpha}+a_{0}^{\alpha\dagger}).
\end{equation}
This gives
\begin{equation}
{\hat X}_{\xi_{i}}=\sqrt{2}\left[V_{0}^{\xi_{i},1}\sqrt{{2\over b}}X_{0}^{1}+\sum_{n=1}^{\infty}V_{2n}^{\xi_{i},1}
{\hat X}_{2n}^{1}+V_{2n-1}^{\xi_{i},2}{\hat P}_{2n-1}^{2}\right],
\end{equation}
\begin{equation}
{\hat Y}_{\xi_{i}}=\sqrt{2}\left[V_{0}^{\xi_{i},2}\sqrt{{2\over b}}X_{0}^{2}+
\sum_{n=1}^{\infty}V_{2n-1}^{\xi_{i},1}{\hat P}_{2n-1}^{1}-iV_{2n}^{\xi_{i},2}{\hat X}_{2n}^{2}\right].
\label{yeq}
\end{equation}

 Since our aim is to obtain the localization in time by making the
 inverse Wick rotation on direction 1, we see that it is $X_{\xi_{i}}$ that contains the time
 coordinate, which we have to compare with the string center of mass time (see below).

\subsection{Projection on the center of mass coordinates}

In order to obtain the open string time profile of the lump solution, we need to
 project it onto the center of mass coordinates of the string. The center
 of mass position operator is given by
\begin{equation}
{\hat x}_{cm,\alpha}={i\over\sqrt{b}}(a_{0,\alpha}-a_{0,\alpha}^{\dagger})
\end{equation}
 and its eigenstate is
\begin{equation}
|X_{CM}\rangle={\sqrt{2\Delta\over\pi b}}\;e^{-{1\over b}x_\a x^\a-{2\over\sqrt{b}}ix_\a a^{\a\,\dagger}_{0}
+{1\over 2}a_{0,\alpha}^{\dagger}a^{\a\, \dagger}_{0}}|\Omega_{\theta,b}\rangle.
\end{equation}

One can project the lump on this state to obtain the center of mass time profile.
However, for  reasons that will be clear later, we will first project on
the $Y_{i}$ momenta,

$$|\Lambda\rangle=\bra{Y_1,Y_2}S\rangle={{\cal N}\over \pi\sqrt{[1+ t_d(\eta_{1})][1+ t_d(\eta_{2})]}}{\rm exp}{1\over 2}
\left[{ t_d(\eta_{1})-1\over t_d(\eta_{1})+1}Y_{1}^{2}+{ t_d(\eta_{2})-1\over t_d(\eta_{2})+1}Y_{2}^{2}\right]$$
\begin{equation}
\times{\rm exp}-{1\over 2}\left[e_{\xi_{i}}^{\dagger}e_{\xi_{i}}^{\dagger}t_d(\eta_{i})+
\int_{-\infty}^{\infty}dk a_{i}^{\dagger}(k)a_{i+1}^{\dagger}(-k)t_c(k)\right]
|\Omega_{e}\rangle \otimes |\Omega_{c}\rangle \otimes |S_{||}\rangle.
\label{lumpdy}
\end{equation}
Where we have used the notation
\begin{equation}
|\Omega_{e}\rangle=\prod_{i=1}^{2}|\Omega_{e_{i}}\rangle,~~~~~~~~~~~ |\Omega_{c}\rangle=\prod_{i=1}^{2} \prod_{k}|\Omega_{i}(k)\rangle.
\end{equation}
Taking equation (\ref{twdefo}) and the corresponding relations for the continuous spectrum oscillators,
equation (\ref{lumpdy}) can be rewritten as
$$|\Lambda\rangle={{\cal N}\over \pi\sqrt{[1+ t_d(\eta_{1})][1+ t_d(\eta_{2})]}}{\rm exp}{1\over 2}
\left[{ t_d(\eta_{1})-1\over t_d(\eta_{1})+1}Y_{1}^{2}+{ t_d(\eta_{2})-1\over t_d(\eta_{2})+1}Y_{2}^{2}\right]$$
\begin{equation}
\times{\rm exp}\left[-{1\over 2}a_{0,\alpha}^{\dagger}{\hat S}_{00}^{\alpha\beta}a_{0,\beta}^{\dagger}-
a_{0,\alpha}^{\dagger}S_{0}^{\alpha}-{1\over 2}a_{n,\alpha}^{\dagger}
{\hat S}_{nm}^{\alpha\beta}a_{m,\beta}^{\dagger}\right]|{\hat \Omega}_{b,\theta}\rangle \otimes |S_{||}\rangle,
\end{equation}
where $|{\hat \Omega}_{b,\theta}\rangle=|\Omega_{e}\rangle \otimes |\Omega_{c}\rangle$ and
\begin{equation}
{\hat S}_{00}^{\alpha\beta}=\sum_{i=1}^{2}V_{0}^{(\xi^{+}_{i})\alpha}
{\bar V}_{0}^{(\xi^{+}_{i})\beta}t_d(\eta_{i})+\int_{-\infty}^{\infty}dkt(k)V_{0}^{i,\alpha}(k){\bar V}_{0}^{i,\beta}(k)
\end{equation}
\begin{equation}
S_{0}^{\alpha}=\sum_{i=1}^{2}\sum_{n=1}\left[V_{0}^{(\xi^{+}_{i})\alpha}
{\bar V}_{n}^{(\xi^{+}_{i})\beta}t_d(\eta_{i})+\int_{-\infty}^{\infty}dkt(k)V_{0}^{i,\alpha}(k)
{\bar V}_{n}^{i,\beta}(k)\right]a_{n,\beta}^{\dagger}={\hat S}_{0n}^{\alpha\beta}a_{n,\beta}^{\dagger}
\end{equation}
\begin{equation}
{\hat S}_{nm}^{\alpha\beta}=\sum_{i=1}^{2}(-1)^{n}V_{n}^{(\xi^{+}_{i})\alpha}
{\bar V}_{m}^{(\xi^{+}_{i})\beta}t_d(\eta_{i})+\int_{-\infty}^{\infty}dkt(k)(-1)^{n}V_{n}^{i,\alpha}(k){\bar V}_{m}^{i,\beta}(k)
\end{equation}
with $ V_{N}^{(\xi^{+}_{i})\alpha}$ being the twist even combination of the discrete eigenstates, see appendix B.
Now let's project onto the center of mass coordinates
$$\langle X_{CM}|\Lambda\rangle={{\cal N}\over \pi\sqrt{[1+ t_d(\eta_{1})][1+ t_d(\eta_{2})]}}
{\rm exp}{1\over 2}\left[{ t_d(\eta_{1})-1\over t_d(\eta_{1})+1}Y_{1}^{2}+{ t_d(\eta_{2})-1\over t_d(\eta_{2})+1}Y_{2}^{2}\right]$$
$$\times \sqrt{\frac{2 \Delta}{\pi b}}{1\over \sqrt{[1+s_{1}][1+s_{2}]}}{\rm exp}{1\over b}
\left[{s_{1}-1\over s_{1}+1}x_{1}x^{1}+{s_{2}-1\over s_{2}+1}x_{2}x^{2}+2i\sqrt{b}
\left({S_{0,1}x^{1}\over 1+s_{1}}+{S_{0,2}x^{2}\over 1+s_{2}}\right)\right]$$
\begin{equation}
\times {\rm exp}\left[-{1\over 2}a_{n,\beta}^{\dagger}\left({\hat S}_{nm}^{\alpha\beta}-{{\hat S}^{\alpha}_{n0,1}
{\hat S}^{1\beta}_{0m}\over 1+s_{1}}-{{\hat S}^{\alpha}_{n0,2}
{\hat S}^{2\beta}_{0m}\over 1+s_{2}}\right)a_{n,\beta}^{\dagger}\right]|0\rangle\otimes|S_{||}\rangle\label{lumpxcm}
\end{equation}
where
$$s_{1}=2\Delta[g_d^{2}(\eta_{1},\eta_2)t_d(\eta_{1})+g_d^{2}(\eta_{2},\eta_1)t_d(\eta_{2})]+
\Delta\int_{-\infty}^{\infty}dk~t_c(k)[g_c^{2}(k)+g_c^{2}(-k)),$$
\begin{equation}
s_{2}=\Delta\int_{-\infty}^{\infty}dk~t_c(k)[g_c^{2}(k)+g_c^{2}(-k)],
\end{equation}
$t_c(k)=-e^{-\pi|k|/2}$ is the eigenvalue of ${\cal T}$ in the continuous spectrum, see
Appendix A for the definition of the remaining terms which enter in the last two equations.
The  inverse Wick-rotation along direction 1 of (\ref{lumpxcm}) should give us a
time-localized solution. It depends on two parameters,  $b$ and $ a$, which can  be
expressed in terms of  ($\eta_{1}, \eta_{2}$), through the eigenvalues equations (\ref{eigend}) .
Let's now take a
look at every term in this solution and analyze it for different values of the such parameters.

 In the Wick-rotated solution, to get time-localization, the term $-\frac {1}{b}\frac{s_{1}-1}{s_{1}+1}$
 should be negative. We cannot achieve this using the conventional lump, since in this case $-1<s_{1}<1$.
 To correct this, as anticipated, we need to invert one or two discrete
 eigenvalues, ($t_d(\eta_{1})$ or/and $t_d(\eta_{2})$). In this case one can easily show that  $1<s_{1}<\infty$ and we get
 the desired behaviour. Given the possibility of inverting one or two eigenvalues, it might seem that
  there is some arbitrariness in our procedure. Actually there is none, since the  cancelation
  of the potentially divergent terms when  $b \to \infty$ (see below),
   requires the inversion of only one eigenvalue. In addition, time localization in small $b$ regime
    requires the inversion of the eigenvalue of $\cal T$ corresponding to the greater between
    $\eta_{1}$ and $\eta_{2}$ ($\eta_{2}$ in our conventions).
 From now on  we will then consider a solution in which $t_d(\eta_{2})$ is inverted, i.e. $t_d(\eta_{2})\to t_d^{-1}(\eta_{2})$.

Next,  look at the term  ${s_{2}-1\over s_{2}+1}x_{2}x^{2}$. Due to the $\bra{Y_1,Y_2}$ projection, it gets a contribution
 only from the continuous spectrum, which is always negative and in the range $(-1,0)$.
As a result, this second term is always negative and gives localization in the transverse  space
direction.

Now we would like to point out some facts about the two parameters
on which our solution depends. In  \cite{BMST}, it has been  pointed out
 that the inverse of the parameter $b$, for large $b$, plays the role of Sen's
 $\tilde\lambda$ near zero. Here again we can repeat the same argument. Note however that
in taking $b$ to infinity we should keep $a$ vanishing, see (\ref{aBb}), since we cannot overcome the
critical value $|B|_c=\frac{1}{2\pi}$.
For this reason the result of taking $b \to \infty$ is
 insensible of the value of the $E$-field, making this limit completely commutative.

As it is justified in Appendix C, the proper way to send $b$ to $\infty $ is to take
 $\eta_{1}\approx\eta_{2}\to\infty$ keeping $\eta_{1}<\eta_{2}$.
 In this case one can easily see that
\begin{equation}
s_{1}\approx \eta_{2}{\rm log}\eta_{1}\eta_{2}+t_c(k_{0}\approx 0),~~~~~~~~~~~~~ s_{2}\approx t_c(k_{0}\approx 0)
\end{equation}
with $k_{0}$ as defined in Appendix C. Note that $t_c(k_{0}\approx
0)=-1$. This is so because the $E$--field cannot scale to infinity
due to existence of critical value.
 Then it follows
\begin{equation}
\lim_{\eta_{1},\eta_{2}\to\infty}{s_{1}-1\over s_{1}+1}
=1,~~~~~~~~~~ \lim_{\eta_{1},\eta_{2}\to\infty}{s_{2}-1\over
s_{2}+1}= -\infty
\end{equation}
As justified in Appendix C.1.2, in this limit  ${\hat
S}^{\alpha\beta(c)}_{n0}=0$ so that the oscillating term in
(\ref{lumpxcm}) receives a contribution only from the discrete
part. It is also pointed out that the discrete contribution
vanishes except for $\alpha=\beta=1$, which is the only non
trivial contribution to the oscillating term. Moreover, we have
\begin{equation}
\lim_{\eta_{1},\eta_{2}\to\infty}{\Delta{\hat S}^{11}_{n0}\over
s_{1}+1}=(-1)^{n} \lim_{\eta_{1},\eta_{2}\to\infty}{1\over
2\sqrt{{\rm log}\eta_{1}\eta_{2}}}=0,
\end{equation}
 Therefore, the oscillating term in (\ref{lumpxcm}) vanishes when $b\to\infty$.

Now let's consider the non-zero mode terms, i.e, the  last line in
(\ref{lumpxcm}). In the $b\to\infty$ limit it is clear that
$V_{n}^{(\xi_{i}^{+})\alpha}$ vanishes for $\alpha=2$ and $ n\ge
1$. Therefore, the contribution of the discrete spectrum to ${\hat
S}_{nm}^{\alpha\beta}$ is zero for  $\alpha$ or $\beta=2$ and $n,
m\ge 1$. However, for $\alpha=\beta=1$ this is not true and there
are potentially divergent contributions from the discrete
spectrum. We are now going to show that  these divergences cancel
and the expression
\begin{equation}
{\check S}_{nm}^{11}= {\hat S}_{nm}^{11(c)}+{\hat S}_{nm}^{11(d)}-{{\hat S}^{1(d)}_{n0,1}{\hat S}^{11(d)}_{0m}\over 1+s_{1}}.
\label{lumpnz}
\end{equation}
is finite when $b\to\infty$.\\
 To this end we notice that, inverting only $t_d(\eta_{2})$
 but taking both $\eta_{1}$ and $\eta_{2}$ to infinity,
the  different terms which enter in the above expression have the
following behaviours
$$1+s_{1}\approx \Delta t_d^{-1}(\eta_{2})V_{0}^{(\xi_{2}^{+}),1}{\bar V}_{0}^{(\xi_{2}^{+}),1},$$
$${\hat S}^{1(d)}_{n0,1}\approx \Delta(-1)^{n}t_d^{-1}(\eta_{2})V_{n}^{(\xi_{2}^{+}),1}{\bar V}_{0}^{(\xi_{2}^{+}),1},$$
$${\hat S}^{11(d)}_{n0}\approx (-1)^{n}t_d^{-1}(\eta_{2})V_{n}^{(\xi_{2}^{+}),1}{\bar V}_{0}^{(\xi_{2}^{+}),1},$$
\begin{equation}
{\hat S}^{11(d)}_{nm}\approx (-1)^{n}t_d^{-1}(\eta_{2})V_{n}^{(\xi_{2}^{+}),1}{\bar V}_{m}^{(\xi_{2}^{+}),1},
\end{equation}
Note that $t_d^{-1}(\eta_{2})=e^{|\eta_{2}|}$  gives a divergent
contribution  as $\eta_2\to\infty$. However, using these results
in eq.(\ref{lumpnz}), it is easy to see that the divergent terms
cancel and we are left with ${\check S}_{nm}^{11}= {\hat
S}_{nm}^{11(c)}.$ This, combined with the fact that for $\alpha=2$
or $\beta=2$ we have  ${\hat S}_{NM}^{\alpha\beta(d)}=0$,  leads
us to the conclusion that ${\check S}_{nm}^{\alpha\beta}= {\hat
S}_{nm}^{\alpha\beta(c)}+O\left({1\over b}\right)$. It is also
verified in Appendix C that ${\hat S}_{nm}^{11(c)}={\hat
S}_{nm}^{22(c)}=S_{nm}$.

This also show that is not possible to  invert both the
  discrete eigenvalues and obtain the same cancelation.
Indeed,  if we invert
    both, the term ${\hat S}^{1(d)}_{n0,1}{\hat S}^{11(d)}_{0m}$
    contains mixed terms like
    $[t^{-1}(\eta_{1})V_{n}^{(\xi_{1}^{+}),1}{\bar V}_{0}^{(\xi_{1}^{+}),1}]
    [t^{-1}(\eta_{2})V_{n}^{(\xi_{2}^{+}),1}{\bar V}_{0}^{(\xi_{2}^{+}),1}]$,
     for which we cannot find a counter term in ${\hat S}_{nm}^{11(d)}$ to cancel it.
     As a result we will not be able to get a regular time and space localized solution,
     since these terms diverge in the limit $\eta_{1}, \eta_{2}\to\infty$.

After all these remarks,
 we can write the space-time localized solution in the
 $b\to\infty$ limit as
\begin{equation}
\lim_{b \to \infty}\langle X_{CM}|\Lambda\rangle_{Wick}= N(Y_{1},
Y_{2})\lim_{b \to \infty}e^{-{\Delta\over
b}(x^{0})^{2}}e^{-\epsilon(b)(x^{2})^{2}}|S\rangle
\end{equation}
where $|S\rangle$ is the space-time independent VSFT solution (the
sliver). Note that time dependence completely disappears in this
limit . A remark is in order for the quantity $\epsilon(b)$ this
number is given by, see (\ref{lumpxcm})
\begin{equation}
\epsilon(b)=\frac\Delta b \frac{s_2-1}{s_2+1}
\end{equation}
a numerical analysis shows that this becomes vanishing as
$b\to\infty$. One can indeed easily check (numerically) that the
$\frac1b$ correction to ${s_2(b)-1\over s_2(b)+1} $ diverges. This
in turn implies that the loss of time dependence is accompanied by
loss of transverse space dependence, giving a resulting zero
momentum state (the D25--sliver). Therefore, taking $b$ to
infinity is like sitting at the original unstable vacuum (the
D25--brane), which is the same situation as setting Sen's
$\tilde\lambda$ to zero.

 Another remark we would like to make is about small $b$ limit, which we can get by taking
 $\eta_{1}\to 0$ and keeping $\eta_{2}$ finite.
Given that  the large $b$ limit corresponds to Sen's
$\tilde\lambda$ near zero (i.e it
  represents the unstable vacuum), it is natural to think that the small $b$ limit
  corresponds to $\tilde\lambda$ near $1\over 2$ (or the stable vacuum). As a matter
  of fact, taking this limit of $b$ one gets the 0 state, which is also obtained
  in the $x_{0}\to\infty$ limit and corresponds to the stable vacuum to which the D-brane decays.
  This can be seen by noting that, in this case, $V_{0}^{\alpha}(k)\to 0$,
whereas $V_{n}^{\alpha}(k)$ for $n\ge 1$ have a finite
nonvanishing limit. As a result $s_{1}$ do not get a contribution
from the continuous spectrum and  $s_{2}=0$. Then, it follows
\begin{equation}
 -{\Delta\over b}{s_{1}-1\over s_{1}+1}\approx -{\Delta\over \eta_{1}}
 \left(\left|\frac{s_{1}-1}{s_{1}+1}\right|\right),
 ~~~~~~~~~{\Delta\over b}{s_{2}-1\over s_{2}+1}\approx -{\Delta\over\eta_{1}}
\end{equation}
where we have used ($b\approx\eta_{1}$), ($s_{1}\approx
1+O(\sqrt\eta_{1})$)  in the limit $\eta_{1}\to 0$ and $\eta_{2}$
finite. These are results one can easily obtain from appendix C.2.
For $\Delta\ne 0$ both of these terms gives a negative infinity in
the exponent and suppress everything in front to give us the 0
state which corresponds to the stable vacuum. However, the case
$\Delta=0$ should be handled with care. In this case, one can send
$\Delta$ and $\eta_{1}$ to zero, in such a way that the ratio
${\Delta\over\eta_{1}}$ remains finite. As a result the time
dependence will be lost while the solution is still space
localized. One should compare this with the time independent
solution obtained when we send Sen's $\tilde\lambda$  to ${1\over
2}$ and, at the same time, tune the  $E$--field  to its critical
value, \cite{Mukho}, obtaining a static fundamental strings
background.

\newsection{Conclusions}

In this work we have shown that the $E$--field deformed star
algebra still presents time localized idempotents, which are a
 generalization of their $E=0$ cousins.

We have analyzed in detail the case of a longitudinal $E$--field
(D25--brane decay)  and of a transverse one (D24--brane decay).

In both cases time localization is achieved by working in the Wick
rotated euclidean theory with an unconventional lump having
inverted discrete eigenvalues of the Neumann matrix  with respect
to the conventional one. In particular the decaying D--24 brane
solution has the nice feature of being localized in time and on
the 25-th spatial direction (although for the extreme case
$b\to\infty$ the transverse space profile becomes flat). This is a
consequence of the fact that we first projected the unconventional
lump on the twist odd momenta of the discrete spectrum and then on
the center of mass coordinates, haven't we done this we would have
lost space localization in the 24--th direction.

Our solution depends on two parameter, $b$ and $E$. After
projecting on the center of mass time coordinate we have shown
that in the extreme cases $b=0$ and $b=\infty$ time dependence is
lost and we are left respectively with the 0 state (VSFT vacuum)
and the sliver state (VSFT D--brane), confirming an earlier
hypothesis, \cite{BMST}, that, although a finite value of $b$ can
be formally changed by a gauge transformation, \cite{RSZ2}, these
two cases, at least qualitatively, seems to reproduce the
$\tilde\lambda=\frac12$ and $\tilde\lambda=0$ of Sen's BCFT,
\cite{Senroll}. The $E$--field is another free parameter in the
range $0\leq E\leq E_c=\frac{1}{2\pi\alpha'}$. For $E$ going to
its critical value, time dependence is lost and we get a flat (non
zero) profile in time (effective tensionless regime), this result
still persists in the $b\to0$ limit if we keep the ratio
$\frac{\Delta}{b}$ finite. This case should correspond to the
tachyon vacuum with a background of fundamental strings, prevented
to decay by their polarization due to critical electric field. Of
course this is a very indirect way to see these fundamental
strings arising in a classical solution of VSFT, a more direct
construction of such objects is given in \cite{lorianofund}.

What we think is  pressing, at the moment, is some workable
way to define the energy momentum tensor for time dependent (V)SFT
solutions, possibly along the lines of \cite{Sen2}. It would allow to extract physical informations, first
of all their energy. This in turn  would prove very useful in
understanding how energy is conserved, while open string degrees
of freedom are suppressed (except the $E$--field charged
fundamental strings) and hence how open/closed duality is implemented, \cite{Senroll, KMS, LLM, GIR}.

\begin{center}
{\bf Acknowledgments}
\end{center}
We thank L. Bonora for collaboration and useful discussions at
various stages of the work. C.M. thanks G. Bonelli and M.
Salizzoni for discussions. This research was supported by the
Italian MIUR under the program ``Teoria dei Campi, Superstringhe e
Gravit\`a'' and by CAPES-Brasil, as far as R.J.S.S. is concerned.

\section*{Appendices}

\appendix
\newsection{Spectroscopy of the Neumann matrices with B field}
In this appendix we present the computation  of the eigenstates
and eigenvalues of the Neumann matrix ${\cal
X}^{\alpha}_{\;\;\beta}$ in the presence of $B$--field, along
the line of \cite{belov1}. A similar analysis was carried out
in \cite{Feng}, but with no reference to the correct normalization
of continuous and discrete eigenvectors; moreover the discrete
eigenvectors presented in the first of \cite{Feng} does not reproduce the known
ones when $B\to 0$. Since the discrete spectrum is of crucial
importance for our purposes we re--derive completely the whole
spectroscopy taking care of the correct normalization of continuous and discrete eigenvalues, as in  \cite{belov1}.
To avoid the degeneracy of the
diagonal Neumann coefficient ${\mathcal X}^{\alpha}_{\;\;\beta}$,
we consider the unitary matrices $C'{\mathcal
U}^{\alpha}_{\;\;\beta}$ and ${\mathcal
U}^{\alpha}_{\;\;\beta}C'$, which are related to ${\cal
X}^{\alpha}_{\;\;\beta}$ as follows \cite{BMS}

\begin{equation}
\left({\mathcal X}^{\alpha}_{\;\;\beta}\right)_{NM}={1\over
3}\left(\delta^{\alpha}_{\;\;\beta}+C'{\mathcal U}^{\alpha}_{\;\;\beta}+C'{\mathcal
{\bar U}}^{\alpha}_{\;\;\beta}\right)_{NM}.
\end{equation}
The matrix $(C'{\mathcal U}^{\alpha}_{\;\;\beta})_{NM}$ can be written explicitly as

\begin{eqnarray}
 C'{\mathcal U}& =& \left( \begin{array}{cccc}
1-3bK & 2{\sqrt {3}b}Ka &
3{\sqrt {b}}K\langle W| & -2{\sqrt {3b}}Ka\langle W|\\
-2{\sqrt {3}b}Ka & 1-3bK & 2{\sqrt
{3b}}Ka\langle W| &
3{\sqrt {b}}K\langle W|\\
3{\sqrt {b}}K|W\rangle & -2{\sqrt {3b}}Ka|W\rangle & CU-3K|W\rangle\langle W| &
2{\sqrt 3}Ka|W\rangle\langle W|\\
2{\sqrt {3b}}Ka|W\rangle & 3{\sqrt {b}}K|W\rangle & -2{\sqrt
3}Ka|W\rangle\langle W| & CU-3K|W\rangle\langle W|\end{array}
\right)\0
\end{eqnarray}
where, see \cite{BMST}

\begin{equation}
|W\rangle =-\sqrt{2}(| {\it v}_{e}\rangle
+i| {\it v}_{o}\rangle),~~~~~~~~~~~~~K={A^{-1}\over 4a^{2}+3}.
\end{equation}
$CU$ is the non-zero mode analog of $C'{\cal U}$ without $B$
field. We recall that, \cite{BMS},

\begin{equation}
C'{\mathcal {\bar U}}={\mathcal {\tilde U}}C',
\end{equation}
where tilde means
transposition with respect to  $\alpha, \beta$ indices.\

Our aim is to solve the
eigenvalue equation

\begin{eqnarray}
C'{\mathcal U}|\Psi\rangle=\xi|\Psi\rangle, ~~~~~~~~~~~~~~~~
|\Psi\rangle= \left( \begin{array}{c}
g_{1}  \\
g_{2}  \\
|\Lambda_{1}\rangle\\
 |\Lambda_{2}\rangle\end{array} \right),
\label{eigeneq}
\end{eqnarray}

which splits into

\begin{equation}
\langle W|\Lambda_{1}\rangle={A\over \sqrt{b}}[\xi-1+{b\over
A}]g_{1}+{2Aa\over \sqrt{3b}}(\xi-1)g_{2}
\label{scaeq1}
\end{equation}

\begin{equation}
\langle W|\Lambda_{2}\rangle={A\over \sqrt{b}}[\xi-1+{b\over
A}]g_{2}-{2Aa\over \sqrt{3b}}(\xi-1)g_{1}
\label{scaeq2}
\end{equation}

\begin{equation}
(CU-\xi)|\Lambda_{1}\rangle={\sqrt {1\over
b}}g_{1}(\xi-1)|W\rangle
\label{vecteq1}
\end{equation}

\begin{equation}
(CU-\xi)|\Lambda_{2}\rangle={\sqrt {1\over
b}}g_{2}(\xi-1)|W\rangle
\label{vecteq2}
\end{equation}

We know, \cite{RSZ5},  that $CU$ has a continuous  spectrum and the
solution of (\ref{eigeneq}) depends on whether the eigenvalue $\xi$
is in the continuous spectrum of $CU$ or not. So we
will distinguish  these two different cases and analyze each
of them in detail.

\subsection{Discrete spectrum}

If $\xi$ is not in the spectrum of $CU$, we can invert
$(CU-\xi)$ in equations (\ref{vecteq1}) and (\ref{vecteq2}) to
obtain

\begin{equation}
|\Lambda_{1}\rangle={\sqrt {1\over b}}g_{1}(\xi-1){1\over
(CU-\xi)}|W\rangle
\end{equation}

\begin{equation}
|\Lambda_{2}\rangle={\sqrt {1\over
b}}g_{2}(\xi-1){1\over (CU-\xi)}|W\rangle .
\end{equation}
As we can see the solutions get modified w.r.t. the $B=0$ case,  only via
possible modifications of the eigenvalue $\xi$.  Substitution of these solutions into equations
(\ref{scaeq1}) and (\ref{scaeq2}) gives

\begin{equation}
{\sqrt{1\over 2b}}(\xi-1)\langle W |{1\over
CU-\xi} | W\rangle g_{1}-{A\over{\sqrt {2b}}}\left(\xi-1+{b\over
A}\right )g_{1}-{2aA\over{\sqrt {6b}}}(\xi-1)g_{2}=0
\label{scaeq3}
\end{equation}

\begin{equation}
{\sqrt{1\over 2b}}(\xi-1)\langle W |{1\over
CU-\xi} | W\rangle g_{2}-{A\over{\sqrt {2b}}}\left(\xi-1+{b\over
A}\right )g_{2}+{2aA\over{\sqrt {6b}}}(\xi-1)g_{1}=0
\label{scaeq4}
\end{equation}
The bracket which appears here is the same as the one in \cite{belov1} and is given by

\begin{equation}
\langle W |{1\over CU-\xi} | W\rangle=V_{00}+\frac{\xi+1}{\xi-1}2\Re F(\eta)
\end{equation}
where

\begin{equation}
F(\eta)=\psi\left(\frac{1}{2}+\frac{\eta}{2\pi i}\right)-\psi\left(\frac{1}{2}\right),~~~~~~~
\xi=-\frac{1}{1-2\rm{cosh}\eta}[2-\rm{cosh}\eta -i\sqrt{3}\rm{sinh}\eta].
\end{equation}
$\psi(x)$ is the logarithmic derivative of the Euler $\Gamma$-function.

Substitution of these in (\ref{scaeq3}) and (\ref{scaeq4}) gives us
\begin{equation}
\left(\Re F(\eta)-{b\over 4}\right)g_{1}-{aA\over{\sqrt 3}}{\xi-1\over
\xi+1}g_{2}=0,~~~~~~
\left(\Re F(\eta)-{b\over 4}\right)g_{2}+{aA\over{\sqrt 3}}{\xi-1\over
\xi+1}g_{1}=0.
\label{syseq}
\end{equation}

This system of equations will have non trivial solutions  for
$g_{2}$ and $g_{1}$ if the determinant of the coefficient matrix
is zero, i.e.

\begin{equation}
{b\over 4}=\Re F(\eta)\pm aA{\rm tanh}{\eta\over 2},
\label{eigend}
\end{equation}

Using equations (\ref{syseq}) we can show that $g_{2}=\pm ig_{1}$.
This is a constraint on $g_{1}$ and $g_{2}$ thus we cannot split
the eigenstates in the two directions, choosing one of the
constants to be zero.  $g_{1}$ is now an overall constant, which
can be chosen real and fixed by normalization completely.

The eigenstates are then

{\bf Case-1 }
\begin{eqnarray}
{b\over 4}=\Re F(\eta)+ aA{\rm tanh}{\eta\over 2},~~~~~~~g_{2}=-ig_{1}=-i g_d(\eta_1,\eta_2)\\
|V^{(\xi_{1})}\rangle= g_d(\eta_{1},\eta_2)\left( \begin{array}{c}
1
\\ -i  \\ {\sqrt {1\over b}}(\xi_{1}-1){1\over CU-\xi_{1}}|W\rangle \\
-i{\sqrt {1\over b}}(\xi_{1}-1){1\over CU-\xi_{1}}|W\rangle \end{array} \right)
\label{eigv1}
\end{eqnarray}

\begin{eqnarray}
|V^{({\bar\xi}_{2})}\rangle= g_d(\eta_{2},\eta_1)\left(
\begin{array}{c} 1
\\ -i  \\ {\sqrt {1\over b}}({\bar\xi}_{2}-1){1\over CU-{\bar\xi}_{2}}|W\rangle \\
-i{\sqrt {1\over b}}({\bar\xi}_{2}-1){1\over CU-{\bar\xi}_{2}}|W\rangle \end{array} \right)
\end{eqnarray}

{\bf Case-2 }
\begin{eqnarray}
{b\over 4}=\Re F(\eta)- aA{\rm tanh}{\eta\over 2},~~~~~~~g_{2}=ig_{1}=i g_d(\eta_2,\eta_1)\\
|V^{(\xi_{2})}\rangle= g_d(\eta_{2},\eta_1)\left( \begin{array}{c}
1
\\ i  \\ {\sqrt {1\over b}}(\xi_{2}-1){1\over CU-\xi_{2}}|W\rangle \\
i{\sqrt {1\over b}}(\xi_{2}-1){1\over CU-\xi_{2}}|W\rangle \end{array} \right)
\end{eqnarray}

\begin{eqnarray}
|V^{({\bar\xi}_{1})}\rangle= g_d(\eta_{1},\eta_2)\left(
\begin{array}{c} 1
\\ i  \\ {\sqrt {1\over b}}({\bar\xi}_{1}-1){1\over CU-{\bar\xi}_{1}}|W\rangle \\
i{\sqrt {1\over b}}({\bar\xi}_{1}-1){1\over CU-{\bar\xi}_{1}}|W\rangle \end{array} \right).\label{eigv2}
\end{eqnarray}
Normalizing them in the following way\footnote{This is the
standard way to normalize eigenvectors of hermitean  matrices}
$$\bar{V}^{\xi_{i}}_{\alpha}V^{\xi_{j},\alpha}=\delta^{ij}$$
$$\bar{V}^{\bar\xi_{i}}_{\alpha}V^{\bar\xi_{j},\alpha}=\delta^{ij}$$
$$\bar{V}^{\bar\xi}_{\alpha}V^{\xi,\alpha}=0$$
we get, use the results of \cite{belov1},
\begin{equation}
|g_d(\eta_{1},\eta_2)|^{2}={1\over
2\Delta}\left[(1-r(\eta_1,\eta_2))+r(\eta_1,\eta_2){\rm
sinh}\eta_{1} {\partial\over\partial\eta_{1}}[{\rm Log}\Re
F(\eta_{1})]\right)^{-1},
\end{equation}
where
\begin{equation}
r(\eta_1,\eta_2)=\Re F(\eta_{1})\frac{{\rm tanh}({\eta_{1}\over
2})+{\rm tanh}({\eta_{2}\over 2})}{\Re F(\eta_{2}){\rm
tanh}({\eta_{1}\over 2})+\Re F(\eta_{1}){\rm tanh}({\eta_{2}\over
2})}.
\end{equation}
It is important to notice that $V^{(\xi_{1})}$  and
$V^{({\bar\xi}_{1})}$ are degenerate eigenstates of $\mathcal X$,
and the same holds for $V^{(\xi_{2})}$ and $V^{({\bar\xi}_{2})}$.

\subsection{Continuous spectrum}

If $\xi$ is in the  continuous spectrum of $CU$ ($\xi=\nu(k)$,
\cite{belov1}), we cannot invert the operator ($CU-\xi$). Thus, in
this case, the solution of (\ref{vecteq1}) and (\ref{vecteq2}) is
\begin{equation}
|\Lambda_{1}\rangle=A_{1}(k)|k\rangle+{1\over\sqrt {b}}g_{1}(\nu(k)-1)\wp{1\over (CU-\nu(k))}|W\rangle
\end{equation}
\begin{equation}
|\Lambda_{2}\rangle=A_{2}(k)|k\rangle +{1\over\sqrt
{b}}g_{2}(\nu(k)-1)\wp{1\over CU-\nu(k)}|W\rangle.
\end{equation}
Where $\wp$ is the principal value, \cite{belov1}. Using these in
(\ref{scaeq1}) and (\ref{scaeq2}), we get
\begin{equation}
A_{1}(k)=g_{1}{\sqrt {2\over b}}k\left(\Re F_{c}(k)-{b\over 4}\right)-
{\sqrt{2}Aa\over{\sqrt {3b}}}k\left({\nu(k)-1\over \nu(k)+1}\right)g_{2}
\end{equation}
\begin{equation}
A_{2}(k)=g_{2}{\sqrt {2\over b}}k\left(\Re F_{c}(k)-{b\over
4}\right)+{\sqrt{2}Aa\over{\sqrt {3b}}}k\left({\nu(k)-1\over \nu(k)+1}\right)g_{1}
\end{equation}
Note that in this case $g_{1}$ and $g_{2}$ are completely
free and we can choose them to construct two linearly independent orthogonal vectors as follows \\
{\bf Case-1 }~~~~$g_{2}=ig_{1}=ig_c(k)$
\begin{eqnarray}
V^{1}(k)= g_{c}(k)\left( \begin{array}{c}
1 \\
i  \\
P(k)|k\rangle+{1\over\sqrt {b}}(\nu(k)-1)\wp{1\over
CU-\nu(k)}|W\rangle-iH(k,a)|k\rangle  \\
 iP(k)|k\rangle+i{1\over\sqrt {b}}(\nu(k)-1)\wp{1\over
CU-\nu(k)}|W\rangle+H(k,a)|k\rangle \end{array} \right)
\end{eqnarray}
{\bf Case-2 }~~~~$g_{2}=-ig_{1}=-ig_c(-k)$
\begin{eqnarray}
V^{2}(k)= g_{c}(-k)\left( \begin{array}{c}
1 \\
-i  \\
P(k)|k\rangle+{1\over\sqrt {b}}(\nu(k)-1)\wp{1\over
CU-\nu(k)}|W\rangle+iH(k,a)|k\rangle  \\
 -iP(k)|k\rangle-i{1\over\sqrt {b}}(\nu(k)-1)\wp{1\over
CU-\nu(k)}|W\rangle+H(k,a)|k\rangle \end{array} \right),
\end{eqnarray}
where
\begin{equation}
P(k)={\sqrt {2\over b}}k\left(\Re F_{c}(k)-{b\over 4}\right), ~~~~~~~~
H(k,a)={\sqrt{2}Aa\over{\sqrt {3b}}}k\left({\nu(k)-1\over \nu(k)+1}\right).
\end{equation}
Imposing the continuous orthonormality condition
\be
\bar{V}^{i,\a}(k)V^{j}_\a(k')=\delta^{ij}\delta(k-k')
\ee
we get
\begin{equation}
g_{c}(k)=\left[{4\Delta\over b}N(k)\left(4+k^{2}\left(\Re
F_{c}(k)-{b\over 4}-{Aa\over{\rm tanh}{\pi
k\over4}}\right)^{2}\right)\right]^{-1/2}
\end{equation}
Sending $k\to-k$ we get the right degeneracy for $\mathcal X$.

\newsection{Diagonalization of the 3-string vertex and the Lump state}

 We can express the oscillators $a_{N,\alpha}^{(r)}$,
 appearing in the 3-string vertex (\ref{3vert}), in terms
 of the oscillators of the diagonal basis as
\begin{equation}
a^{(r)}_{N,\alpha}=\sum_{i=1}^{2}\left(a^{(r)}_{\xi_{i}}\bar{V}^{(\xi_{i})}_{N,\alpha}+
a^{(r)}_{{\bar\xi}_{i}}\bar{V}^{({\bar\xi}_{i})}_{N,\alpha}+\int_{-\infty}^{\infty}dka^{(r)}_{i}(k)\bar{V}^{(i)}_{N,\alpha}(k)\right)
\end{equation}

\begin{equation}
a^{(r)\dagger}_{N,\alpha}=\sum_{i=1}^{2}\left(a^{(r)\dagger}_{\xi_{i}}
V^{(\xi_{i})}_{N,\alpha}+a^{(r)\dagger}_{{\bar\xi}_{i}} V^{({\bar\xi}_{i})}_{N,\alpha}+
\int_{-\infty}^{\infty}dka^{(r)\dagger}_{i}(k) V^{(i)}_{N,\alpha}(k)\right).
\end{equation}

 Using these oscillators and the fact that $\tau{\bar V}=V$
 ($\tau^{\alpha}_{\beta}=\left( \begin{array} {cc} 1 & 0 \\ 0 & -1 \end{array}\right)$),
 we can rewrite the 3-string vertex as

$$|V^{m}_{3}\rangle=N_{m}{\rm exp}[-{1\over 2}\sum_{r,s}\sum_{i=1}^{2}
\left(a^{(r)\dagger}_{\xi_{i}}{\bar V}^{(\xi_{i})}_{N,\alpha}+a^{(r)\dagger}_{{\bar\xi}_{i}}{\bar V}^{({\bar\xi}_{i})}_{N,\alpha}+
\int_{-\infty}^{\infty}dka^{(r)\dagger}_{i}(k){\bar V}^{(i)}_{N,\alpha}(k)\right)(\tau C'{\mathcal X})^{\alpha(rs)}_{\beta,NM}$$
\begin{equation}\label{opty}
\times \sum_{j=1}^{2}\left(a^{(s)\dagger}_{\xi_{j}} V^{(\xi_{j})\beta}_{M}+a^{(s)\dagger}_{{\bar\xi}_{j}}
V^{({\bar\xi}_{j})\beta}_{M}+\int_{-\infty}^{\infty}dk'a^{(s)\dagger}_{j}(k') V^{(j)\beta}_{M}(k')\right)]|\Omega_{b,\theta}\rangle
\end{equation}

The twist operator $\tau C'$ acts on the eigenstates of the
discrete and continous spectra as follows
\begin{equation}
\tau C'V^{(\xi_{i})}=V^{({\bar\xi}_{i})}~~~~~~~~~\tau C'V^{(i)}(k)=V^{(i+1)}(-k),
\end{equation}
where $V^{3}(k)$ is identified with $V^{1}(k)$. Then (\ref{opty})
becomes

$$|V^{m}_{3}\rangle=N_{m}{\rm exp}\left[-{1\over 2}\sum_{r,s}
\sum_{i=1}^{2}\left(a^{(r)\dagger}_{\xi_{i}}\mu^{rs}({\bar\xi}_{i})a^{(s)\dagger}_{{\bar\xi}_{i}}+a^{(r)\dagger}_{{\bar\xi}_{i}}
\mu^{rs}(\xi_{i})a^{(s)\dagger}_{\xi_{i}}\right.\right.+$$
\begin{equation}\label{poiu}
\left. \left.+\int_{-\infty}^{\infty}dka^{(r)\dagger}_{i}(k)\mu^{rs}(-k)a^{(s)\dagger}_{i+1}(-k)\right)\right]|\Omega_{b}\rangle.
\end{equation}

In order to write this in an exact diagonal form, we need to
introduce oscillators with definite $\tau$--twist parity
\begin{equation}\label{deftwist}
e^{r}_{\xi_{i}}={a^{r}_{\xi_{i}}+a^{r}_{{\bar\xi}_{i}}\over\sqrt{2}}
={a^{r}_{\xi_{i}}+\tau
C'a^{r}_{\xi_{i}}\over\sqrt{2}},~~~~~~~o^{r}_{\xi_{i}}=-i{a^{r}_{\xi_{i}}-a^{r}_{{\bar\xi}_{i}}\over\sqrt{2}}=-i{a^{r}_{\xi_{i}}-\tau
C'a^{r}_{\xi_{i}}\over\sqrt{2}}
\end{equation}

$$e^{r}_{i}(k)={a^{r}_{i}(k)+a^{r}_{i+1}(-k)\over\sqrt{2}}={a^{r}_{i}(k)+\tau C'a^{r}_{i}(k)\over\sqrt{2}}$$
\begin{equation}
\end{equation}
$$o^{r}_{i}(k)=-i{a^{r}_{i}(k)-a^{r}_{i+1}(-k)\over\sqrt{2}}=-i{a^{r}_{i}(k)-\tau C'a^{r}_{i}(k)\over\sqrt{2}}$$

These oscillators have the following BPZ conjugation property
\begin{equation}
{\rm bpz}~o_{i}=-o^{\dagger}_{i}~~~~~~~~~{\rm
bpz}~e_{i}=-e^{\dagger}_{i}\;,
\end{equation}
and  satisfy the commutation relations
$$[e_{\xi_{i}},e^{\dagger}_{\xi_{j}}]=\delta_{ij},~~~~~~~[o_{\xi_{i}},o^{\dagger}_{\xi_{j}}]=\delta_{ij},$$
\begin{equation}
[e_{i}(k),e^{\dagger}_{j}(k')]=\delta_{ij}\delta(k-k'),~~~~~~[o_{i}(k),o^{\dagger}_{j}(k')]=\delta_{ij}\delta(k-k'),
\end{equation}
with all the other commutators vanishing. Using  them  into
(\ref{poiu}) we finally obtain

$$|V^{m}_{3}\rangle= N_{m}{\rm exp}\left[-{1\over 4}\sum_{r,s}\sum_{i=1}^{2}
\left( \left[\mu^{rs}(\xi_{i})+\mu^{rs}({\bar\xi}_{i})\right]
\left(e^{(r)\dagger}_{\xi_{i}}e^{(s)\dagger}_{\xi_{i}}+o^{(r)\dagger}_{\xi_{i}}o^{(s)\dagger}_{\xi_{i}}\right)\right. \right.$$
$$\left. \left.-i \left[\mu^{rs}(\xi_{i})-\mu^{rs}({\bar\xi}_{i})\right]
\left(o^{(r)\dagger}_{\xi_{i}}e^{(s)\dagger}_{\xi_{i}}-e^{(r)\dagger}_{\xi_{i}}o^{(s)\dagger}_{\xi_{i}}\right)\right. \right.$$
$$\left. \left.-\int_{-\infty}^{\infty}dk\mu^{rs}(k)\left(e^{(r)\dagger}_{i}(k)
e^{(s)\dagger}_{i}(k)+o^{(r)\dagger}_{i}(k)o^{(s)\dagger}_{i}(k)\right)\right. \right.$$
\begin{equation}
\left. \left.-i\int_{-\infty}^{\infty}dk\mu^{rs}(k)\left(e^{(r)\dagger}_{i}(k)
o^{(s)\dagger}_{i}(k)-o^{(r)\dagger}_{i}(k)e^{(s)\dagger}_{i}(k)\right)\right) \right]|\Omega_{b,\theta}\rangle
\end{equation}

This gives the diagonal representation of the 3-string interaction vertex.
 The same procedure gives the following diagonal representation of the transverse part of the Lump

$$|S_{\bot}\rangle= \frac{A^{2}(3+4a^{2})}{\sqrt{2\pi b^{3}}(\Det G)^{\frac{1}{4}}}
 \Det({\cal I}-{\cal X})^{\frac{1}{2}}\Det({\cal I}+{\cal T})^{\frac{1}{2}}
{\rm exp}\left(-{1\over
2}\sum_{i=1}^{2}\left[t_d(\eta_{i})\left(e^{\dagger}_{\xi_{i}}
e^{\dagger}_{\xi_{i}}+o^{\dagger}_{\xi_{i}}o^{\dagger}_{\xi_{i}}\right)\right.
\right.+$$
\begin{equation}
\left. \left.+{1\over 2}\int_{-\infty}^{\infty}dk
t_c(k)\left(e^{\dagger}_{i}(k)
e^{\dagger}_{i}(k)+o^{\dagger}_{i}(k)o^{\dagger}_{i}(k)\right)\right]\right)|\Omega_{b,\theta}\rangle\label{lumpd}
\end{equation}

\newsection{Asymptotic behaviours}

In Section 3.3, we have analyzed our solution in the large and
small limits of the parameter $b$. In this appendix we compute the
relevant matrix elements in these asymptotic regimes.

\subsection{The $b\rightarrow \infty$ Limit}

 From
(\ref{eigend}), we can write
\begin{equation}
|a|=\frac{\Re F(\eta_{2})-\Re F(\eta_{1})}{[V_{00}+2\Re F(\eta_{2})]{\rm tanh}({\eta_{1}\over 2})+
[V_{00}+2\Re F(\eta_{1})]{\rm tanh}({\eta_{2}\over 2})}
\label{aeq}
\end{equation}
\begin{equation}
\frac{b}{4}=\frac{\Re F(\eta_{2}){\rm tanh}({\eta_{1}\over 2})+\Re F(\eta_{1}){\rm tanh}({\eta_{2}\over 2})}
{{\rm tanh}({\eta_{1}\over 2})+{\rm tanh}({\eta_{2}\over 2})}
\label{beq}
\end{equation}
where we take, by definition, $\eta_{2} > \eta_{1}>0$. There are
two ways of taking $b \rightarrow \infty$\\ {\bf i) $\eta_{2}
\rightarrow \infty$ ; $\eta_{1}$ fixed }

In this limit we can see that
\begin{equation}
\frac{b}{4} \approx \left(\frac{{\rm tanh}({\eta_{1}\over 2})}
{1+{\rm tanh}({\eta_{1}\over 2})}\right){\rm
log}(\eta_{2}),~~~~~~~~~~~a \approx \frac{1}{2{\rm
tanh}({\eta_{1}\over 2})} > \frac{1}{2}.
\end{equation}
{\bf ii) $\eta_{2} \rightarrow \infty$ ; $\eta_{1} \rightarrow \infty$ }

We can parameterize $\eta_{2}=\eta^{y}$, $\eta_{1}=\eta^{x}$ and
then take $\eta \rightarrow \infty$, while keeping $y > x$. We
then obtain
\begin{equation}
\frac{b}{4} \approx \frac{1}{2} (y+x){\rm log}(\eta),~~~~~~~~~a \approx \frac{1}{2} \frac{y-x}{y+x} < \frac{1}{2}
\end{equation}
We will be concerned with this second regime as it is the one
connected to $a=0$, which is a condition arising from the
existence of the critical value  for the $E$--field, when
$b\to\infty$. In this second limit it can be easily seen that the
discrete eigenvectors  have the following behaviour
$$V_{0}^{\xi_{i},1}=V_{0}^{\bar{\xi}_{i},1}\approx {1\over\sqrt{2\Delta}}e^{-\eta_{i}/2}\sqrt{\eta_{i}{\rm Log}\eta_{1}\eta_{2}},$$
$$V_{0}^{\xi_{i},2}=-V_{0}^{\bar{\xi}_{i},2}\approx (-1)^{i}{i\over\sqrt{2\Delta}}e^{-\eta_{i}/2}
\sqrt{\eta_{i}{\rm Log}\eta_{1}\eta_{2}},$$ and
\begin{equation}
V_{n}^{\xi_{i},\alpha}\approx-{V_{0}^{\xi_{i},\alpha}\over\sqrt{{\rm Log}\eta_{1}\eta_{2}}},~~~~~~~~~~~
V_{n}^{{\bar\xi}_{i},\alpha}\approx-{V_{0}^{{\bar\xi}_{i},\alpha}\over\sqrt{{\rm Log}\eta_{1}\eta_{2}}}.
\end{equation}
For the continuous spectrum the situation is more complicated and
getting this limits is not easy. However, it is possible to
calculate the limit of $(V_{0}^{i,\alpha}(k))^{2}$, which is
enough for our purposes. We have
\begin{equation}
 (V_{0}^{1,1}(k))^{2}=\left[{4\Delta\over b}N(k)\left(4+k^{2}\left(\Re F_{c}(k)-{b\over
4}-{Aa\over{\rm tanh}{\pi k\over4}}\right)^{2}\right)\right]^{-1}.
\end{equation}
When $b\to\infty$ this expression vanishes every where except at
\begin{equation} k_{0}\approx -{4\over\pi}{\rm arctanh}(2a)
\end{equation}

where it diverges. Expanding  around $k_0$ one  easily gets
$$(V_{0}^{1,1}(k))^{2}\approx\Delta^{-1}{4a\over k_{0}(1-4a^{2})N(k_{0})}{{\bar b}\over\pi(1+(k-k_{0})^{2}{\bar b}^{2}}$$
where $${\bar b}={k_{0}\pi(1-4a^{2})\over 64a}b.$$
Now taking the $b\rightarrow\infty$ limit one obtains
\begin{equation}
(V_{0}^{1,1}(k))^{2}\approx{1\over 2\Delta}\delta(k-k_{0}).
\end{equation}
 Following the same procedure one can also show that
\begin{equation}
(V_{0}^{2,1}(k))^{2}\approx{1\over 2\Delta}\delta(k+k_{0})
\end{equation}
 remember that
\begin{equation}
|V_{0}^{1,2}(k)|^{2}=(V_{0}^{1,1}(k))^{2},~~~~~~~~~|V_{0}^{2,2}(k)|^{2}=(V_{0}^{2,1}(k))^{2}.
\end{equation}

The non zero components,  $V_{m}^{i,\alpha}(k)$, can be expressed
in terms of a generating function. For instance, the generating
function for  $V_{m}^{1,1}(k)$ is given by
\begin{equation}
F^{(k)}(z)=A_{1}(k){\it f}^{(k)}(z)-
\frac{(1-\nu(k))V_{0}^{1,1}(k)}{\sqrt{b}}B(k,z)
\end{equation}
where
$$
A_{1}(k)=V_{0}^{1,1}(k)\sqrt{\frac{2}{b}}k\left(\Re F_{c}(k)-\frac{b}{4}-{Aa\over{\rm tanh}({\pi k\over 4})}\right),
$$
$$
B(k,z)=\frac{2}{1-\nu(k)}\left[\Re F_{c}(k)+\frac{\pi}{2\sqrt{3}}
\frac{\nu(k)-1}{\nu(k)+1} + \frac{2i}{k} + {\rm log}(iz)- 2i\it{f}^{(k)}(z)
\right]$$
\begin{equation}
~~~~~~~~~~~~~~~~+{2\over 1-\nu(k)}\left[\Phi(e^{-4i{\rm
arctan}(z)},1,1+\frac{k}{4i}) e^{-4i{\rm arctan}(z)}e^{-k{\rm
arctan}(z)}\right]
\end{equation}
where $\Phi$ is the LerchPhi function and $f^{(k)}$ is the
generating function for the spectrum of the Neumann matrix without
zero modes, \cite{RSZ5}. Inverting this equation we can write
$V_{m}^{1,1}(k)$ as
\begin{equation}
V_{m}^{1,1}(k)=A_{1}(k)\frac{\sqrt{m}}{2\pi i}\oint dz
\frac{{\it f}^{(k)}(z)}{z^{m+1}}-\frac{(1-\nu(k))V_{0}^{1,1}(k)}{\sqrt{b}}
\frac{\sqrt{m}}{2\pi i}\oint dz \frac{B(k,z)}{z^{m+1}}
\end{equation}
With the same procedure one can also write
\begin{equation}
V_{m}^{2,1}(k)=A'_{1}(k)\frac{\sqrt{m}}{2\pi i}\oint dz
\frac{{\it f}^{(k)}(z)}{z^{m+1}}-\frac{(1-\nu(k))V_{0}^{2,1}(k)}{\sqrt{b}}
\frac{\sqrt{m}}{2\pi i}\oint dz \frac{B(k,z)}{z^{m+1}}
\end{equation}
with
\begin{equation}
A'_{1}(k)=V_{0}^{2,1}(k)\sqrt{\frac{2}{b}}k\left(\Re F_{c}(k)-\frac{b}{4}+{Aa\over{\rm tanh}({\pi k\over 4})}\right).
\end{equation}
 The other vectors are related to these ones as
\begin{equation}
V_{n}^{1,2}(k)=iV_{n}^{1,1}(k),~~~~~~~~~~~~~V_{n}^{2,2}(k)=-iV_{n}^{2,1}(k)
\end{equation}

\subsubsection{Limit of ${\hat S}_{mn}^{\alpha\beta(c)}$}

With all these results at hand we can now calculate the continuous
spectrum contribution to the non zero mode matrix elements in the
limits under consideration. Recalling that spectrum of the Neumann
matrix without zero modes is given by
\begin{equation}
v_{m}^{(k)}=\frac{\sqrt{m}}{2\pi i}\oint dz
\frac{{\it f}^{(k)}(z)}{z^{m+1}}
\end{equation} we can write
\begin{equation}
{\hat S}^{11(c)}_{nm}=\int_{-\infty}^{\infty} dk \ t_c(k)(-1)^{n}\left[ V_{n}^{1,1}(k)
 {\bar V}_{m}^{1,1}(k)+ V_{n}^{2,1}(k) {\bar V}_{m}^{2,1}(k)\right]
\end{equation} as
$${\hat S}_{mn}^{11(c)}=\int_{-\infty}^{\infty} dk \ t_c(k)(-1)^{m}[A_{1}(k)
A_{1}(k)v_{m}^{(k)}v_{n}^{(k)}
-A_{1}(k) V_{0}^{1,1}(k) v_{m}^{(k)}(1-\bar{\nu}(k))\tilde{B}_{n}(k)
\frac{1}{\sqrt{b}}$$
$$-A_{1}(k) V_{0}^{1,1}(k) v_{n}^{(k)}(1-\nu(k))\tilde{B}_{m}(k)
\frac{1}{\sqrt{b}}+ (V_{0}^{1,1}(k))^{2}(1-\bar{\nu}(k))(1-\nu(k))
\tilde{B}_{m}(k)\tilde{B}_{n}(k)\frac{1}{b}]$$
\begin{equation}
~~~~~~~~~~~+\left[A_{1}(k)\rightarrow A'_{1}(k), V_{0}^{1,1}(k)\rightarrow V_{0}^{2,1}(k)\right]
\end{equation}

where
\begin{equation}
\tilde{B}_{m}(k)=\frac{\sqrt{m}}{2 \pi i} \oint dz
\frac{B(k,z)}{z^{m+1}}.
\end{equation}
 Note that if the indices are separated
by comma then the first index is  the label of the vector and the
second is the space time index, otherwise both are space time
indices. Now we want to calculate each term in the above
expression in the limit when $b\rightarrow \infty$. To this end we
notice the following
$$
\lim_{b \to \infty}A_{1}(k)A_{1}(k)=\lim_{b \to \infty}(V_{0}^{1,1}(k))^{2}
\left(\frac{2k^{2}}{b}\right)\left(\Re F_{c}(k)-\frac{b}{4}-{Aa\over{\rm tanh}({\pi k\over 4})}\right)^{2}$$
\begin{equation}
~~~~~~~~~~~=\lim_{x \to -\infty}\left(\frac{k^{2}}{2\Delta N(k)}\right)
\frac{x^{2}}{4+k^{2}x^{2}}=
\left(\frac{k^{2}}{2\Delta N(k)}\right)\frac{1}{k^{2}}=\frac{1}{2\Delta N(k)}
\end{equation}

where $x=\left(\Re F_{c}(k)-\frac{b}{4}-{Aa\over{\rm tanh}({\pi
k\over 4})}\right)$. The other terms are zero in the limit
because, either they contain  term like $(k-k_{0})\delta(k-k_{0})$
in the integral or they are of order ${1\over b}$. Therefore, we
are left with
\begin{equation}
\lim_{b \to \infty} {\hat S}_{mn}^{11(c)}=\lim_{b \to \infty} {\hat S}_{mn}^{22(c)}=
\Delta^{-1}S_{mn},~~~{\rm where}~~~S_{nm}=-\int_{-\infty}^{\infty} {dk \ t_c(k)\over N(k)}v_{n}^{(k)}v_{m}^{(-k)}
\end{equation}
and
\begin{equation}
 \lim_{b \to \infty} {\hat S}_{mn}^{21(c)}=\lim_{b \to \infty} {\hat S}_{mn}^{12(c)}=0,
\end{equation}
which is the sliver in each direction with corrections of order
$\frac{1}{b}$.

\subsubsection{Limit of ${\hat S}_{0m}^{\alpha\beta(c)}$}

In this section we would like to justify  that the contribution
from the continuous spectrum to ${\hat S}_{0m}^{\alpha\beta}$  is
zero in the limit. This can be computed the same way as before
since we have
\begin{equation}
\lim_{b \to \infty}{\hat S}_{0m}^{\alpha\beta(c)}= \lim_{b \to
\infty} \sum_{i=1}^{2}\int_{-\infty}^{\infty} dk \ t_c(k)
V_{0}^{i,\alpha}(k)V_{m}^{i,\beta}(k).
\end{equation}
For instance, lets calculate ${\hat S}_{0m}^{11(c)}$ which is
given by
$$
{\hat S}_{0m}^{11(c)}=\lim_{b \to \infty}\int_{-\infty}^{\infty}dk
\ t_c(k) v_{m}^{(k)}\sqrt{2\over b}k\left((V_{0}^{1,1}(k))^{2}
\left[\Re F_{c}(k)-{b\over 4}-{Aa\over{\rm tanh}({\pi k\over 4})}\right]\right.
$$
\begin{equation}
+\left. (V_{0}^{2,1}(k))^{2}\left[\Re F_{c}(k)-{b\over 4}+{Aa\over{\rm tanh}({\pi k\over 4})}\right]\right)+O({1\over\sqrt b})
\end{equation}
We have already verified that $\lim_{b \to
\infty}(V_{0}^{i,\alpha}(k))^{2}\approx {1\over 2}\delta (k\pm
k_{0})$. This will allow us to expand the terms in square brackets
about the points $\pm k_{0}$ to get
 $${\hat S}_{0m}^{11(c)}={1\over\Delta}\lim_{b \to \infty}\int_{-\infty}^{\infty}dk
\ t_c(k) v_{m}^{(k)}\left[{1\over 2}\delta (k- k_{0})
\sqrt{b}(k-k_{0})k_{0}\pi\left({1-4a^{2}\over 32a}\right)\right.$$
\begin{equation}
+\left.{1\over 2}\delta (k+ k_{0})\sqrt{b}(k+k_{0})(-k_{0})\pi\left({1-4a^{2}\over 32a}\right)\right]+O({1\over\sqrt b}).
\end{equation}
 Due to the presence of the delta functions the terms $(k\pm k_{0})\sqrt{b}$ are both finite in the $b\to\infty$ limit.
 As a matter of this fact we can safely do the integrals first and take the limits later. Since the integrals vanishes we note that
\begin{equation}
{\hat S}_{0m}^{11(c)}\approx 0.
\end{equation}
 Similar steps show that all the remaining terms of ${\hat S}_{0m}^{\alpha\beta(c)}$ are also zero.

\subsection{The $b\rightarrow 0$ Limit}

As it was mentioned before this limit can be obtained by taking $\eta_{1}\to 0$. In this limit it is not hard to see that
\begin{equation}
b\approx 2\frac{\Re F(\eta_{2})}{{\rm tanh}({\eta_{2}\over 2})}\eta_{1}
\end{equation}
\begin{equation}
g_d(\eta_{1},\eta_{2})\approx{1\over\sqrt{2\Delta}}\left(1- {{\rm tanh}({\eta_{2}\over 2})\over 2\Re F(\eta_{2})}\eta_{1}\right)
\end{equation}
\begin{equation}
g_d(\eta_{2},\eta_{2})\approx {1\over\sqrt{2\Delta}}\left[2{\rm tanh}({\eta_{2}\over 2})\left({\rm sinh}\eta_{2}
{\partial\over\partial\eta_{2}}[{\rm Log}\Re F(\eta_{2})]-1\right)\right]^{-1/2}\sqrt{\eta_{1}}.
\end{equation}
One can use these results and equations (\ref{eigv1}) through
(\ref{eigv2}) to write down $V_{0}^{\xi_{i},\alpha}$,
$V_{0}^{{\bar\xi}_{i},\alpha}$ $V_{n}^{\xi_{i},\alpha}$ and
$V_{n}^{{\bar\xi}_{i},\alpha}$ as
$$V_{0}^{\xi_{1},1}=V_{0}^{\bar{\xi}_{1},1}\approx {1\over\sqrt{2\Delta}}
\left(1- {{\rm tanh}({\eta_{2}\over 2})\over 2\Re F(\eta_{2})}\eta_{1}\right) ,$$
\begin{equation}
V_{0}^{\xi_{1},2}=-V_{0}^{\bar{\xi}_{1},2}\approx
-i{1\over\sqrt{2\Delta}} \left(1- {{\rm tanh}({\eta_{2}\over
2})\over 2\Re F(\eta_{2})}\eta_{1}\right),
\end{equation}
$$V_{0}^{\xi_{2},1}=V_{0}^{\bar{\xi}_{2},1}\approx {1\over\sqrt{2\Delta}}
\left[2{\rm tanh}({\eta_{2}\over 2})\left({\rm sinh}\eta_{2}
{\partial\over\partial\eta_{2}}[{\rm Log}\Re F(\eta_{2})]-1\right)\right]^{-1/2}\sqrt{\eta_{1}}$$
\begin{equation}
V_{0}^{\xi_{2},2}=-V_{0}^{\bar{\xi}_{2},2}\approx
i{1\over\sqrt{2\Delta}} \left[2{\rm tanh}({\eta_{2}\over
2})\left({\rm sinh}\eta_{2} {\partial\over\partial\eta_{2}}[{\rm
Log}\Re F(\eta_{2})]-1\right)\right]^{-1/2}\sqrt{\eta_{1}},
\end{equation}
and
$$
V_{n}^{\xi_{1},\alpha}=\pm V_{n}^{{\bar\xi}_{1},\alpha} \approx
\sqrt{\eta_{1}},$$
\begin{equation}
V_{n}^{\xi_{2},\alpha}=\pm V_{n}^{{\bar\xi}_{2},\alpha} \approx
{1\over\sqrt{2\Delta}}\left[2{\rm tanh}({\eta_{2}\over
2})\left({\rm sinh}\eta_{2} {\partial\over\partial\eta_{2}}[{\rm
Log}\Re F(\eta_{2})]-1\right)\right]^{-1/2}f(\eta_{2}).
\end{equation}
The $f$ is a regular function of $\eta_{2}$.
On the other hand
\begin{equation}
g_c(k)\approx 0.
\end{equation}
This shows all $V_{0}^{i,\alpha}(k)$ are zero,  whereas
$V_{m}^{i,\alpha}(k)$ are finite and $b$ independent to the
leading order. These results are extensively used in section 3.3
to calculate quantities like $s_{1}$, $s_{2}$ in the $b\to0$
limit.

\end{document}